%% file: sigproc-sp.tex
\documentclass{acm_proc_article-sp}
\usepackage[english]{babel}
\usepackage{amsmath,epsfig}
\usepackage{amsfonts}
\usepackage{amssymb}
\usepackage{graphicx,enumerate,multicol}
\usepackage[ruled]{algorithm2e}
\usepackage{aas_macros}

\include{macros}

\begin{document}

\title{Linear inverse problems with noise: primal and primal-dual splitting\footnote{Submitted to NCMIP 2011 on the 02/20/11.}}

\numberofauthors{3} 

\author{
\alignauthor
Fran{\c c}ois-Xavier Dup\'e\\
       \affaddr{AIM UMR CNRS - CEA}\\
       \affaddr{91191 Gif-sur-Yvette, France}\\
      \email{francois-xavier.dupe@cea.fr}
\alignauthor
Jalal M. Fadili
       \affaddr{GREYC-ENSICAEN-Universit\'e de Caen}\\
       \affaddr{14050 Caen, France}\\
      \email{jfadili@greyc.ensicaen.fr}
\alignauthor
Jean-Luc Starck\\
       \affaddr{AIM UMR CNRS - CEA}\\
       \affaddr{91191 Gif-sur-Yvette, France}\\
      \email{jean-luc.starck@cea.fr}
}
\date{20 February 2011}

\maketitle
\begin{abstract}
  In this paper, we propose two algorithms for solving linear inverse problems when the observations are corrupted by
  noise. A proper data fidelity term (log-likelihood) is introduced to reflect the statistics of the noise
  (e.g. Gaussian, Poisson).  On the other hand, as a prior, the images to restore are assumed to be positive and
  sparsely represented in a dictionary of waveforms. Piecing together the data fidelity and the prior terms, the
  solution to the inverse problem is cast as the minimization of a non-smooth convex functional. We establish the
  well-posedness of the optimization problem, characterize the corresponding minimizers, and solve it by means of primal
  and primal-dual proximal splitting algorithms originating from the field of non-smooth convex optimization
  theory. Experimental results on deconvolution, inpainting and denoising with some comparison to prior methods are also
  reported.
\end{abstract}




\section{Introduction}
\label{sec:intro}


A lot of works have already been dedicated to linear inverse problems with Gaussian noise (see \cite{Starck2010} for a
comprehensive review), while linear inverse problems in presence of other kind of noise such as Poisson noise have
attracted less interest, presumably because noises properties are more complicated to handle. Such inverse problems have
however important applications in imaging such as restoration (e.g. deconvolution in medical and astronomical imaging),
or reconstruction (e.g. computerized tomography).


Since the pioneer work for Gaussian noise by \cite{Daubechies2004}, many other methods have appeared for managing linear
inverse problem with sparsity regularization. But they limited to the Gaussian case.  In the context of Poisson linear
inverse problems using sparsity-promoting regularization, a few recent algorithms have been proposed. For example,
\cite{Dupe2009c} stabilize the noise and proposed a family of nested schemes relying upon proximal splitting algorithms
(Forward-Backward and Douglas-Rachford) to solve the corresponding optimization problem. The work of \cite{Chaux2009} is
in the same vein. These methods may be extended to other kind of noise. However, nested algorithms are time-consuming
since they necessitate to sub-iterate. Using the augmented Lagrangian method with the alternating method of multipliers
algorithm (ADMM), which is nothing but the Douglas-Rachford splitting applied to the Fenchel-Rockafellar dual problem,
\cite{Figueiredo2010} presented a deconvolution algorithm with TV and sparsity regularization, and \cite{Bioucas2010} a
denoising algorithm for multiplicative noise. This scheme however necessitates to solve a least-square problem which can
be done explicitly only in some cases.

In this paper, we propose a framework for solving linear inverse problems when the observations are corrupted by
noise. In order to form the data fidelity term, we take the exact likelihood. As a prior, the images to restore are
assumed to be positive and sparsely represented in a dictionary of atoms. The solution to the inverse problem is cast as
the minimization of a non-smooth convex functional, for which we prove well-posedness of the optimization problem,
characterize the corresponding minimizers, and solve them by means of primal and primal-dual proximal splitting
algorithms originating from the realm of non-smooth convex optimization theory. Convergence of the algorithms is also
shown. Experimental results and comparison to other algorithms on deconvolution are finally conducted.


\subsection*{Notation and terminology}
\label{sec:notation}

Let $\Hc$ a real Hilbert space, here a finite dimensional vector subspace of $\mathbb{R}^n$. We denote by $\norm{.}$ the
norm associated with the inner product in $\Hc$, and $\Id$ is the identity operator on $\Hc$. $\norm{.}_p, p \geq 1$ is
the $\ell_p$ norm. $\vx$ and $\va$ are respectively reordered vectors of image samples and transform coefficients. We
denote by $\ri \Cc$ the relative interior of a convex set $\Cc$. A real-valued function $f$ is coercive, if
$\lim_{\norm{\vx} \to +\infty}f\parenth{\vx}=+\infty$, and is proper if its domain is non-empty $\dom f = \{ x\in\Hc
\mid f(x) < +\infty \} \neq \emptyset$. $\Gamma_0(\Hc)$ is the class of all proper lower semicontinuous (lsc) convex
functions from $\Hc$ to $(-\infty,+\infty]$. We denote by $\opnorm{\mathbf{M}}= \max_{\vx \neq 0}
\frac{\norm{\mathbf{M}\vx}}{\norm{\vx}}$ the spectral norm of the linear operator $\mathbf{M}$, and
$\kerm(\mathbf{M}):=\{x \in \Hc: \mathbf{M}x=0, x \neq 0\}$ its kernel.
  

Let $x \in \Hc$ be an $\sqrt{n}\times\sqrt{n}$ image. $x$ can be written as the superposition of elementary atoms
$\varphi_\gamma$ parameterized by $\gamma \in \mathcal{I}$ such that $x = \sum_{\gamma \in \mathcal{I}} \alpha_\gamma
\varphi_\gamma = \Phb \va,\quad \abs{\mathcal{I}} = L, ~ L\ge n$. We denote by $\Phb: \Hc' \to \Hc$ the dictionary
(typically a frame of $\Hc$), whose columns are the atoms all normalized to a unit $\ell_2$-norm

\section{Problem statement}
\label{sec:problem-statement}

Consider the image formation model where an input image of $n$ pixels $\vx$ is indirectly observed through the action of
a bounded linear operator $\Hmb: \Hc \to \Kc$, and contaminated by a noise $\varepsilon$ through a composition operator
$\odot$ (e.g. addition),
\begin{equation}
  \label{eq:2}
  \vy \sim \Hmb \vx \odot \varepsilon~.
\end{equation}
The linear inverse problem at hand is to reconstruct $\vx$ from the observed image $\vy$.

A natural way to attack this problem would be to adopt a maximum a posteriori (MAP) bayesian framework with an
appropriate likelihood function {\textemdash} the distribution of the observed data $\vy$ given an original $\vx$
{\textemdash} reflecting the statistics of the noise. As a prior, the image is supposed to be economically (sparsely)
represented in a pre-chosen dictionary $\Phb$ as measured by a sparsity-promoting penalty $\Psi$ supposed throughout to
be convex but non-smooth, e.g. the $\ell_1$ norm.

\subsection{Gaussian noise case}
\label{sec:gaussian-noise-case}

For Gaussian noise, we consider the following formation model,
\begin{equation}
  \label{eq:21}
  \vy = \Hmb \vx + \varepsilon~,
\end{equation}
where $\varepsilon \sim \mathcal{N}(0,\sigma^2)$.

From the probability density function, the negative log-likelihood writes:
\begin{equation}
  \label{eq:4}
  f_{\mathrm{Gaussian}}\ : \eta \in \mathbb{R}^n \mapsto \norm{\eta - y}^2_2 / (2\sigma^2)~.
\end{equation}


From this function, we can directly derive the following result,
\begin{propo}
  $f_{\mathrm{Gaussian}}$ is a proper, strictly convex and lsc function.
\end{propo}

\subsection{Poisson noise case}
\label{sec:poisson-noise-case}

The observed image is then a discrete collection of counts $\vy=(\vy[i])_{1 \le i \le n}$ which are bounded, i.e. $\vy
\in \ell_{\infty}$. Each count $y[i]$ is a realization of an independent Poisson random variable with a mean $(\Hmb
\vx)_i$. Formally, this writes in a vector form as {\small
\begin{equation}
  \label{eq:22}
  \vy \sim \Pc(\Hmb \vx)~.
\end{equation}
}

From the probability density function of a Poisson random variable, the likelihood writes: 
{\footnotesize
\begin{equation}
  \label{eq:7}
  p(y|x) = \prod_i \frac{((\Hmb x)[i])^{y[i]} \exp\left(-(\Hmb x)[i]\right)}{y[i]!}~.
\end{equation}}
Taking the negative log-likelihood, we arrive at the following data fidelity term:
{\footnotesize
\begin{align}
  \label{eq:9}
  f_{\mathrm{Poisson}}\ &: \eta \in \mathbb{R}^n \mapsto \sum_{i=1}^n f_{\mathrm{p}}(\eta[i]), \\
  \text{if } y[i] > 0,\quad
  f_{\mathrm{p}}(\eta[i]) &=
  \begin{cases}
    -y[i] \log(\eta[i]) + \eta[i] & \text{if } \eta[i] > 0,\\
    +\infty & \text{otherwise,}
  \end{cases} \nonumber \\
  \text{if } y[i] = 0,\quad
  f_{\mathrm{p}}(\eta[i]) &=
  \begin{cases}
    \eta[i] & \text{if } \eta[i] \in [0,+\infty), \\
    +\infty & \text{otherwise.}
  \end{cases} \nonumber
\end{align}
}

Using classical results from convex theory, we can show that,
\begin{propo}
  $f_{\mathrm{Poisson}}$ is a proper, convex and lsc function. $f_{\mathrm{Poisson}}$ is strictly convex if and
  only if $\forall i \in \{1,\ldots,n\}, y[i] \ne 0$.
\end{propo}

\subsection{Multiplicative noise}
\label{sec:multiplicative-noise}

We consider the case without linear operator and as in \cite{Bioucas2010} with a $M$-look full
developed speckle noise,
\begin{equation}
  \label{eq:13}
  y = x \varepsilon,\quad \varepsilon \sim \Gamma(M,1/M)~.
\end{equation}

In order to simplify the problem, the logarithm of the observation is considered, $\log(y) = \log(x) + \log(\varepsilon)
= z + \omega$. And in \cite{Bioucas2010}, the authors proof that the anti log-likelihood yields,
\begin{equation}
  \label{eq:14}
  f_{\mathrm{Multi}}\ : \ \eta \in \mathbb{R}^n \mapsto M \sum_{i=1}^{n} \left(z[i] + \exp(\log(y[i]) - z[i] \right)~.
\end{equation}

Using classical results from convex theory, we can directly derive,
\begin{propo}
  $f_{\mathrm{Multi}}$ is a proper, strictly convex and lsc function.
\end{propo}

\subsection{Optimization problem}
\label{sec:optimization-problem}

Our aim is then to solve the following optimization problems, under a synthesis-type sparsity prior\footnote{Our
  framework and algorithms extend to an analysis-type prior just as well.}, {\footnotesize
\begin{equation}
  \label{eq:11}
  \begin{gathered}
    \tag{$\Pm_{f_1,\eqi,\psi}$} \argmin_{\va\in\Hc'} J(\va), \\
    J\ :\ \va \mapsto {f_1\circ\Hmb\circ\Phb(\va)} + \eqi \Psi(\va) + \imath_{\Cc} \circ\Phb(\va)~.
  \end{gathered}
\end{equation}
} The data fidelity term $f_1$ reflect the noise statistics, the penalty function $\Psi : \va \mapsto \sum_{i=0}^{L}
\psi_i(\va[i])$ is positive, additive, and chosen to enforce sparsity, $\eqi > 0$ is a regularization parameter and
$\imath_{\Cc}$ is the indicator function of the convex set $\Cc$ (e.g. the positive orthant for Poissonian data).

For the rest of the paper, we assume that $f_1$ is a proper, convex and lsc function, i.e. $f_1\in\Gamma_0(\Hc)$. This
is true for many kind of noises including Poisson, Gaussian, Laplacian\ldots (see \cite{Combettes2007b} for others
examples).

\pagebreak
From the objective in \eqref{eq:11}, we get the following,
\begin{propo}
\label{prop:objectives}
  {~}\\ \vspace*{-0.8cm}
  \begin{enumerate}[(i)]
    \setlength{\topsep}{0pt}
    \setlength{\parskip}{0pt}
    \setlength{\itemsep}{0pt}
    \setlength{\partopsep}{0pt}
  \item $f_1$ is a convex functions and so are $f_1 \circ \Hmb$ and $f_1\circ\Hmb\circ\Phb$. 
  \item Suppose that $f_1$ is strictly convex, then $f_1\circ\Hmb\circ\Phb$ remains strictly convex if $\Phb$ is an
    orthobasis and $\kerm(\Hmb) = \emptyset$.
  \item Suppose that $(0,+\infty) \cap \Hmb\left([0,+\infty)\right) \neq \emptyset$. Then $J \in \Gamma_0(\Hc)$.
  \end{enumerate}
\end{propo}

\subsection{Well-posedness of \eqref{eq:11}}
\label{sec:char-solut}

Let $\Mc$ be the set of minimizers of problem \eqref{eq:11}. Suppose that $\Psi$ is coercive. Thus $J$ is
coercive. Therefore, the following holds:
\begin{propo}
  {~} \\ \vspace{-0.8cm}
  \begin{enumerate}[(i)]
    \setlength{\topsep}{0pt}
    \setlength{\parskip}{0pt}
    \setlength{\itemsep}{0pt}
    \setlength{\partopsep}{0pt}
  \item Existence: \eqref{eq:11} has at least one solution, i.e. $\Mc\ne\emptyset$.
  \item Uniqueness: \eqref{eq:11} has a unique solution if $\Psi$ is strictly convex, or under (ii) of Proposition~\ref{prop:objectives}.
  \end{enumerate}
\end{propo}

\section{Iterative Minimization Algorithms}
\label{sec:sparse-iter-deconv}

\subsection{Proximal calculus}
We are now ready to describe the proximal splitting algorithms to solve \eqref{eq:11}. At the heart of the splitting framework is the notion of proximity operator.
\begin{definition}[\cite{Moreau1962}]
  \label{def:1}
  Let $F \in \Gamma_{0}(\Hc)$. Then, for every $x\in\Hc$, the function $y \mapsto F(y) + \norm{x-y}^{2}_2/2$
  achieves its infimum at a unique point denoted by $\prox_{F}x$. The operator $\prox_{F} : \Hc \to \Hc$
  thus defined is the \textit{proximity operator} of $F$.
\end{definition}

Then, the proximity operator of the indicator function of a convex set is merely its orthogonal projector. One important
property of this operator is the separability property:
\begin{lemma}[\cite{Combettes2005}]
  \label{lem:decomp}
  Let $F_k \in \Gamma_0(\Hc),\ k \in \{1,\cdots,K\}$ and let $G : (x_k)_{1\le k\le K} \mapsto \sum_k F_k(x_k)$. Then
  $\prox_{G} = (\prox_{F_k}) _{1 \le k \le K}$.
\end{lemma}

For Gaussian noise, we can easily prove that with $f_1$ as defined in \eqref{eq:4},
\begin{lemma}
  \label{lem:prgaus}
  Let $y$ be the observation,  the proximity operator associated to $f_{\mathrm{Gaussian}}$ (i.e. the Gaussian anti
  log-likelihood) is,
  \begin{equation}
    \label{eq:6}
    \prox_{\beta f_{\mathrm{Gaussian}}} \vx = \frac{\beta y + \sigma^2 x}{\beta + \sigma^2}~.
  \end{equation}
\end{lemma}

The following result can be proved easily by solving the proximal optimization problem in Definition~\ref{def:1} with
$f_1$ as defined in \eqref{eq:9}, see also \cite{Combettes2007a}.
\begin{lemma}
  \label{lem:prpois}
  Let $y$ be the count map (i.e. the observations), the proximity operator associated to $f_{\mathrm{Poisson}}$ (i.e. the Poisson anti
  log-likelihood) is,
    \begin{equation}
      \label{eq:3}
      \prox_{\beta f_{\mathrm{Poisson}}} \vx = \left( 
        \frac{\vx[i] - \beta + \sqrt{(\vx[i] -\beta)^2 + 4\beta \vy[i]}}{2}
      \right)_{1\le i \le n}~.
    \end{equation}
\end{lemma}

As with multiplicative noise $f_{\mathrm{Multi}}$ involves the exponential, we need the W-Lambert
function~\cite{Corless1996} in order to derive a closed form of the proximity operator,
\begin{lemma}
  \label{lem:prwlam}
  Let $y$ be the observations, the proximity operator associated to $f_{\mathrm{Multi}}$ is,
    \begin{equation}
      \label{eq:31}
      \prox_{\beta f_{\mathrm{Multi}}} \vx  = x - \beta M - \mathrm{W}\left(-\beta M \exp(x - \log(y) - \beta M )\right)~,
    \end{equation}
    where $\mathrm{W}$ is the W-Lambert function.
\end{lemma}

We now turn to $\prox_{\eqi\Psi}$ which is given by Lemma~\ref{lem:decomp} and the following result:
\begin{theorem}[\cite{Fadili2006}]
  \label{th:3}
  Suppose that $\forall~ i$: (i) $\psi_i$ is convex even-symmetric, non-negative and non-decreasing on $\mathbb{R}^+$,
  and $\psi_i(0)=0$; (ii) $\psi_i$ is twice differentiable on $\mathbb{R}\setminus \{0\}$; (iii) $\psi_i$ is continuous
  on $\mathbb{R}$, and admits a positive right derivative at zero ${\psi_i^{'}}_+(0) = \lim_{h\to 0^+}
  \frac{\psi_i(h)}{h} > 0$. Then, the proximity operator $\prox_{\delta\psi_i}(\beta) = \hat{\va}(\beta)$ has exactly
  one continuous solution decoupled in each coordinate $\beta[i]$ :
  \begin{equation}
    \label{eq:10}
    \hat{\va}[i] =
    \begin{cases}
      0 & \text{if } \abs{\beta[i]} \le \delta{\psi_i^{'}}_+(0)\\
      \beta_i-\delta\psi_i^{'}(\hat{\va}[i]) & \text{if } \abs{\beta[i]} > \delta{\psi_i^{'}}_+(0)
    \end{cases}
  \end{equation}
\end{theorem}
Among the most popular penalty functions $\psi_i$ satisfying the above requirements, we have $\psi_i(\va[i]) =
\abs{\va[i]}, \forall ~ i$, in which case the associated proximity operator is soft-thresholding, denoted $\mathrm{ST}$
in the sequel.

\subsection{Splitting on the primal problem}
\label{sec:primal-method}

\subsubsection{Splitting for sums of convex functions}
\label{sec:splitt-sums-conv}

Suppose that the objective to be minimized can be expressed as the sum of $K$ functions in $\Gamma_0(\Hc)$, verifying domain qualification conditions:
\begin{equation}
  \label{eq:sum}
  \argmin_{x \in \Hc} ~ \left(F(x) = \sum_{k=1}^K F_k(x)\right)~.
\end{equation}
Proximal splitting methods for solving \eqref{eq:sum} are iterative algorithms which may evaluate the individual proximity
operators $\prox_{F_k}$, supposed to have an explicit convenient structure, but never proximity operators of sums of the $F_k$.

Splitting algorithms have an extensive literature since the 1970's, where the case $K=2$ predominates. Usually,
splitting algorithms handling $K > 2$ have either explicitly or implicitly relied on reduction of \eqref{eq:5} to the
case $K = 2$ in the product space $\Hc^K$. For instance, applying the Douglas-Rachford splitting to the reduced form
produces Spingarn's method, which performs independent proximal steps on each $F_k$, and then computes the next iterate
by essentially averaging the individual proximity operators. The scheme described in \cite{Combettes2008} is very
similar in spirit to Spingarn's method, with some refinements.

\subsubsection{Application to noisy inverse problems}
\label{sec:appl-poiss-noise}

Problem \eqref{eq:11} is amenable to the form \eqref{eq:sum}, by wisely introducing auxiliary variables. As
\eqref{eq:11} involves two linear operators ($\Phb$ and $\Hmb$), we need two of them, that we define as $\vx_1 =
\Phb\va$ and $\vx_2 = \Hmb\vx_1$. The idea is to get rid of the composition of $\Phb$ and $\Hmb$. Let the two linear
operators $\Lmb_1 =[ \Id \quad 0 \quad -\Phb]$ and $\Lmb_2 = [ -\Hmb \quad \Id \quad 0]$. Then, the optimization problem
\eqref{eq:11} can be equivalently written:
\begin{gather}
  \label{eq:1}
  \argmin_{(\vx_1,\vx_2,\va) \in \Hc\times\Kc\times\Hc'} \underbrace{f_1(\vx_2) + \imath_{\Cc}(\vx_1) + \eqi\Psi(\va)}_{G(\vx_1,\vx_2,\va)} + \\
  \imath_{\ker \Lmb_1}(\vx_1,\vx_2,\va) + \imath_{\ker \Lmb_2}(\vx_1,\vx_2,\va)~.
\end{gather}
Notice that in our case $K=3$ by virtue of separability of the proximity operator of $G$ in $x_1$, $x_2$ and $\alpha$;
see Lemma~\ref{lem:decomp}.

\begin{algorithm}[h]
  \noindent{\bf{Parameters:}} The observed image $y$, the dictionary $\Phb$, number of iterations
  $N_{\mathrm{iter}}$, $\mu > 0$ and regularization parameter $\eqi > 0$. \\
  \noindent{\bf{Initialization:}}\\
  $\forall i \in \{1,2,3\},\quad p_{(0,i)} = (0,0,0)^\mathrm{T}$. $z_0 = (0,0,0)^\mathrm{T}$. \\
  \noindent{\bf{Main iteration:}} \\
  \noindent{\bf{For}} $t=0$ {\bf{to}} $N_{\mathrm{iter}}-1$,
  \begin{itemize}
    \setlength{\topsep}{0pt}
    \setlength{\parskip}{0pt}
    \setlength{\itemsep}{0pt}
    \setlength{\partopsep}{0pt}
  \item \underline{Data fidelity} (Lemmas~\ref{lem:prgaus}, \ref{lem:prpois} and \ref{lem:prwlam}): $\xi_{(t,1)}[1] = \prox_{\mu f_1/3}(p_{(t,1)}[1])$.
  \item \underline{Sparsity-penalty} (Lemma~\ref{th:3}): $\xi_{(t,1)}[2] =
    \prox_{\mu \eqi\Psi/3}(p_{(t,1)}[2])$.
  \item \underline{Positivity constraint}: $\xi_{(t,1)}[3] = \Prj_{{\Cc}}(p_{(t,1)}[3])$.
  \item \underline{Auxiliary constraints with $\Lmb_1$ and $\Lmb_2$:} (Lemma~\ref{th:prjli}):
    $\xi_{(t,2)} = \Prj_{{\ker \Lmb_1}}(p_{(t,2)}), \xi_{(t,3)} = \Prj_{{\ker \Lmb_2}}(p_{(t,3)})$.
  \item Average the proximity operators: $\xi_{t} = (\xi_{(t,1)} + \xi_{(t,2)} + \xi_{(t,3)})/3$. 
  \item Choose $\theta_t\in]0,2[$.
  \item Update the components: $\forall i \in \{1,2,3\},\quad p_{(t+1,i)} = p_{(t,i)} + \theta_t (2\xi_{t} - z_{t} - \xi_{(t,i)})$.
  \item Update the coefficients estimate: $z_{t+1} = z_t + \theta_t(\xi_t - z_t)$.
  \end{itemize}
  \noindent{\bf{End main iteration}} \\
  \noindent{\bf{Output:}} Reconstructed image $x^{\star}=z_{N_{\mathrm{iter}}}[0]$.
  \caption{Primal scheme for solving \eqref{eq:11}.}
  \label{algo:deconv}
\end{algorithm}

The proximity operators of $F$ and $\Psi$ are easily accessible through Lemmas~\ref{lem:prgaus}, \ref{lem:prpois},
\ref{lem:prwlam} and \ref{th:3}. The projector onto $\Cc$ is trivial for most of the case (e.g. positive orthant, closed
interval). It remains now to compute the projector on $\ker \Lmb_i$, $i=1,2$, which by well-known linear algebra
arguments, is obtained from the projector onto the image of $\Lmb_i^*$.
\begin{lemma}
  \label{th:prjli}
  The proximity operator associated to $\imath_{\ker \Lmb_i}$ is
  \begin{gather}
    \label{eq:10a}
    \Prj_{\ker \Lmb_i} = \Id - \Lmb_i^* (\Lmb_i \circ \Lmb_i^*)^{-1} \Lmb_i~.
  \end{gather}
\end{lemma}
The inverse in the expression of $\Prj_{\ker \Lmb_1}$ is $(\Id + \Phb\circ\Phi^\mathrm{T})^{-1}$ can be computed
efficiently when $\Phb$ is a tight frame. Similarly, for $\Lmb_2$, the inverse writes $(\Id + \Hmb\circ\Hmb^*)^{-1}$,
and its computation can be done in the domain where $\Hmb$ is diagonal; e.g. Fourier for convolution or pixel domain for
mask.

Finally, the main steps of our primal scheme are summarized in Algorithm~\ref{algo:deconv}. Its convergence is a
corollary of \cite{Combettes2008}[Theorem~3.4].

\begin{propo}
  Let $(z_t)_{t\in\mathbb{N}}$ be a sequence generated by Algorithm~\ref{algo:deconv}. Suppose that
  Proposition~\ref{prop:objectives}-(iii) is verified, and $\sum_{t\in\mathbb{N}} \theta_t(2-\theta_t) = +\infty$.
Then $(z_t)_{t\in\mathbb{N}}$ converges to a (non-strict) global minimizer of \eqref{eq:11}.
\end{propo}

\subsubsection{Splitting on the dual: Primal-dual algorithm}
\label{sec:primal-dual-method}

Our problem \eqref{eq:11} can also be rewritten in the form,
\begin{gather}
  \label{eq:5}
  \argmin_{\va \in \Hc'} F\circ\Kmb(\va) + \eqi\Psi(\va)
\end{gather}
where now $\Kmb = \begin{pmatrix}\ \Hmb\circ\Phb\ \\ \ \Phb \end{pmatrix}$ and $F : (\vx_1,\vx_2) \mapsto f_1(x_1) +
\imath_{\Cc}(x_2)$. Again, one may notice that the proximity operator of $F$ can be directly computed using the
separability in $\vx_1$ and $\vx_2$.

Recently, a primal-dual scheme, which turns to be a pre-conditioned version of ADMM, to minimize objectives of the form
\eqref{eq:5} was proposed in \cite{Chambolle2010}. Transposed to our setting, this scheme gives the steps summarized in
Algorithm~\ref{algo:deconv2}.

Adapting the arguments of \cite{Chambolle2010}, convergence of the sequence $(\va_t)_{t\in\mathbb{N}}$ generated by Algorithm~\ref{algo:deconv2} is ensured.
\begin{propo}
  Suppose that Proposition~\ref{prop:objectives}-(iii) holds. Let $\zeta = \opnorm{\Phb}^2(1+\opnorm{\Hmb}^2)$, choose $\tau > 0$ and $\sigma$
  such that $\sigma\tau\zeta < 1$, and let $(\va_t)_{t\in\mathbb{R}}$ as defined by
  Algorithm~\ref{algo:deconv2}. Then, $(\va)_{t\in\mathbb{N}}$ converges to a (non-strict) global minimizer \eqref{eq:11} at the rate $O(1/t)$ on the restricted duality gap.
\end{propo}

\subsection{Discussion}
\label{sec:discussion}

Algorithm~\ref{algo:deconv} and \ref{algo:deconv2} share some similarities, but exhibit also important differences. For
instance, the primal-dual algorithm enjoys a convergence rate that is not known for the primal algorithm. Furthermore,
the latter necessitates two operator inversions that can only be done efficiently for some $\Phb$ and $\Hmb$, while the
former involves only application of these linear operators and their adjoints. Consequently,
Algorithm~\ref{algo:deconv2} can virtually handle any inverse problem with a bounded linear $\Hmb$. In case where the
inverses can be done efficiently, e.g. deconvolution with a tight frame, both algorithms have comparable computational
burden. In general, if other regularizations/constraints are imposed on the solution, in the form of additional proper
lsc convex terms that would appear in \eqref{eq:11}, both algorithms still apply by introducing wisely chosen auxiliary
variables.

\begin{algorithm}[h]
  \noindent{\bf{Parameters:}} The observed image $y$, the dictionary $\Phb$, number of iterations
  $N_{\mathrm{iter}}$, proximal steps $\sigma > 0$ and $\tau > 0$, and regularization parameter $\eqi > 0$. \\
  \noindent{\bf{Initialization:}}\\
  $\va_0 = \bar{\va}_0 = 0$
  $\xi_0 = \eta_0 = 0$. \\
  \noindent{\bf{Main iteration:}} \\
  \noindent{\bf{For}} $t=0$ {\bf{to}} $N_{\mathrm{iter}}-1$,
  \begin{itemize}
    \setlength{\topsep}{0pt}
    \setlength{\parskip}{0pt}
    \setlength{\itemsep}{0pt}
    \setlength{\partopsep}{0pt}
  \item \underline{Data fidelity} (Lemmas~\ref{lem:prgaus}, \ref{lem:prpois} and \ref{lem:prwlam}): $\xi_{t+1} = (\Id - \sigma \prox_{f_1/\sigma})(\xi_{t}/\sigma + \Hmb\circ\Phb \bar{\va}_{t})$.
  \item \underline{Positivity constraint}: $\eta_{t+1} = (\Id - \sigma\Prj_{{\Cc}})(\eta_{t}/\sigma + \Phb\bar{\va}_{t})$.
  \item \underline{Sparsity-penalty} (Lemma~\ref{th:3}):
    $\va_{t+1} = \prox_{\tau \eqi\Psi} \left(\va_t - \tau\Phb^\mathrm{T}\left(\Hmb^*\xi_{t+1} + \eta_{t+1}\right)\right)$.
  \item Update the coefficients estimate: $\bar{\va}_{t+1} = 2\va_{t+1} - \va_t$
 \end{itemize}
  \noindent{\bf{End main iteration}} \\
  \noindent{\bf{Output:}} Reconstructed image $x^{\star}=\Phb\va_{N_{\mathrm{iter}}}$.
  \caption{Primal-dual scheme for solving \eqref{eq:11}.}
  \label{algo:deconv2}
\end{algorithm}

\section{Experimental results}
\label{sec:results}

\subsection{Deconvolution under Poisson noise}
\label{sec:deconv-under-poiss}

\begin{figure}[htb]
  \centering
  \includegraphics[width=0.4\linewidth]{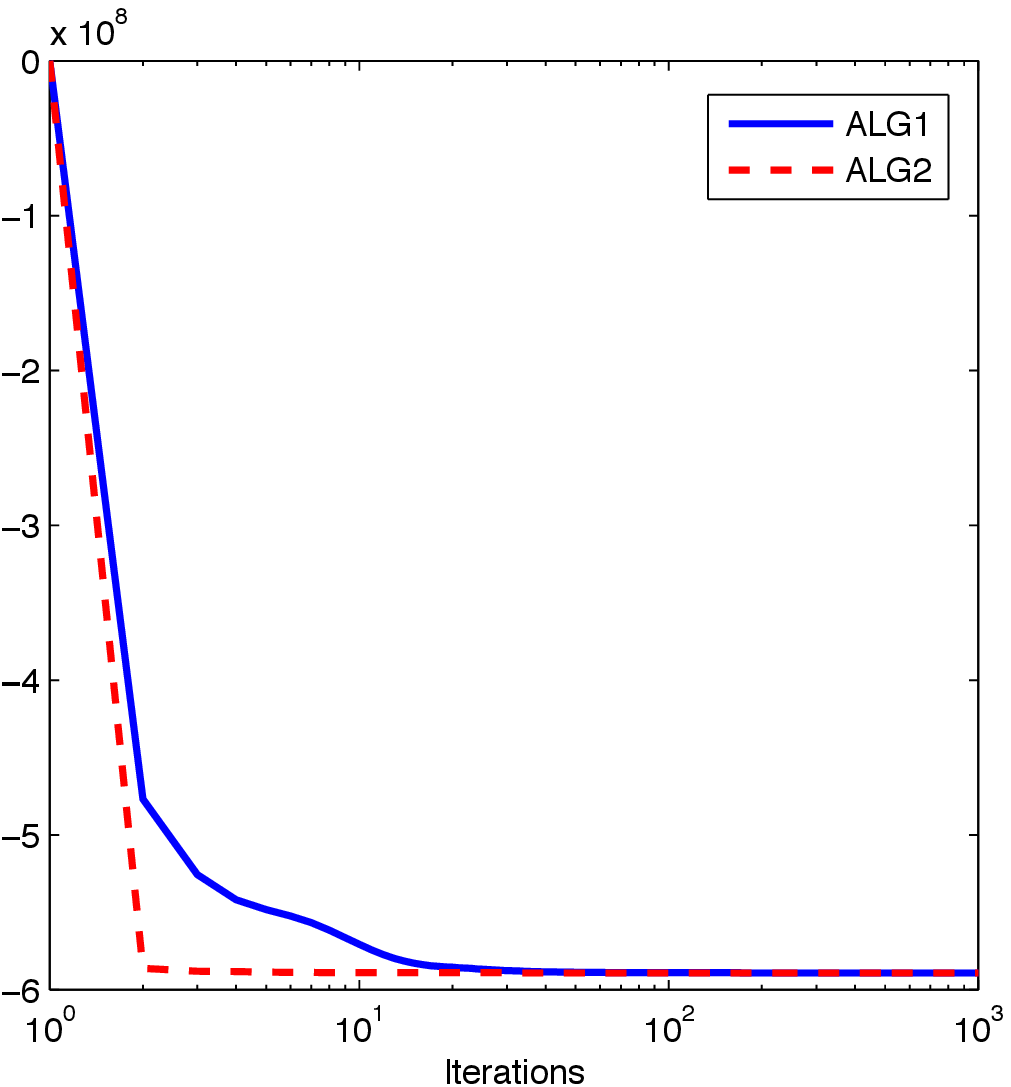}
  \includegraphics[width=0.4\linewidth]{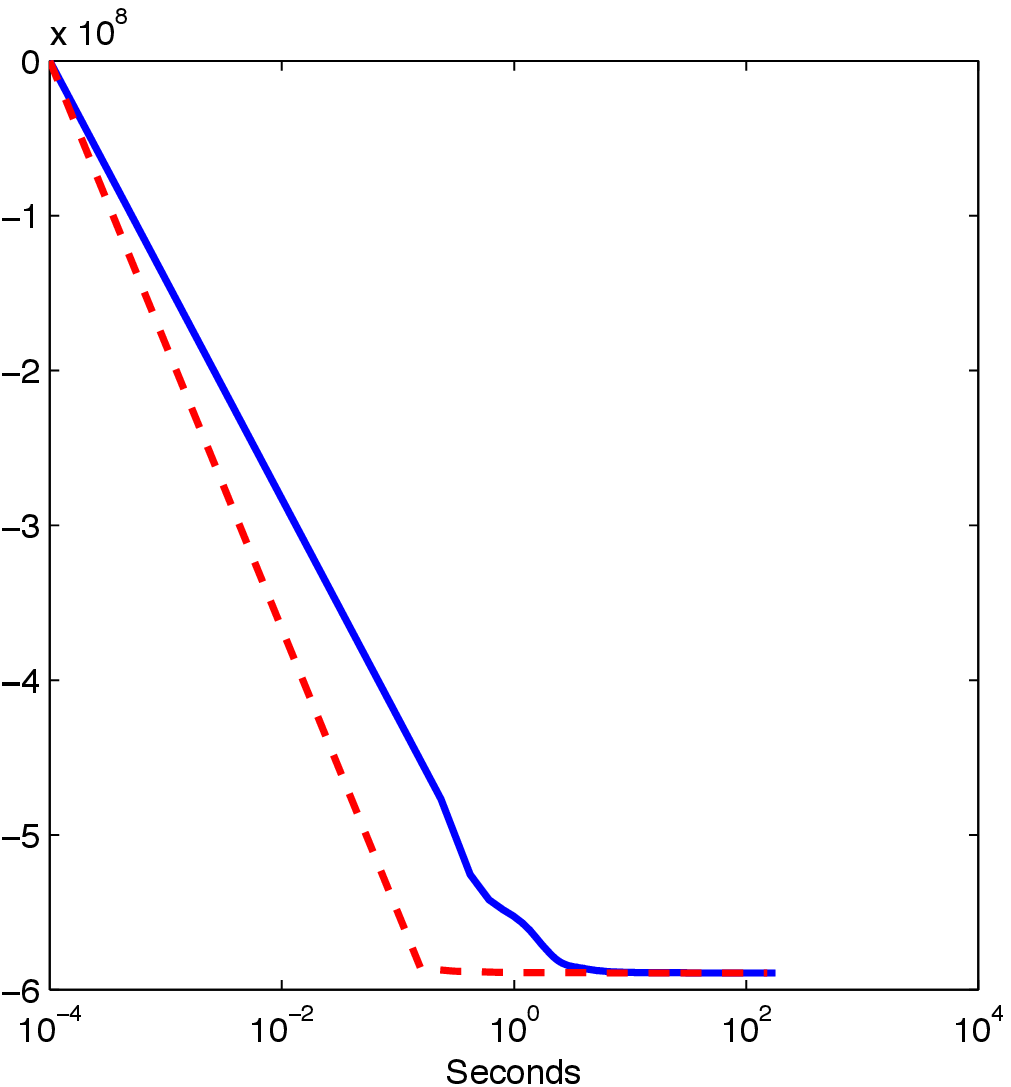}
  \caption{Objective function for deconvolution under Poisson noise in function if iterations (left) and times (right).}
  \label{fig:plot}
\end{figure}

Our algorithms were applied to deconvolution. In all experiments, $\Psi$ was the $\ell_1$-norm. Table~\ref{tab:maesky}
summarizes the mean absolute error (MAE) and the execution times for an astronomical image, where the dictionary
consisted of the wavelet transform and the PSF was that of the Hubble telescope. Our algorithms were compared to
state-of-the-art alternatives in the literature. In summary, flexibility of our framework and the fact that Poisson
noise was handled properly, demonstrate the capabilities of our approach, and allow our algorithms to compare very
favorably with other competitors. The computational burden of our approaches is also among the lowest, typically faster
than the PIDAL algorithm. Fig.~\ref{fig:plot} displays the objective as a function of the iteration number and times (in
s). We can clearly see that Algorithm~2 converges faster than Algorithm~1.


\begin{table}[htb]
\hspace*{-0.5cm}
  \centering
  \scriptsize
  \begin{tabular}{|l||c|c|c|c|}\hline
    & RL-MRS \cite{Starck2006}& RL-TV \cite{Dey2004}& StabG \cite{Dupe2009c} & PIDAL-FS \cite{Figueiredo2010} \\
    \hline MAE & 63.5      & 52.8    & 43  &  43.6  \\\hline
    Times & 230s & 4.3s & 311s & 342s  \\ \hline \cline{1-3}

   &  Alg.~\ref{algo:deconv} & Alg.~\ref{algo:deconv2} \\ \cline{1-3}
   MAE & 46     & 43.6 \\ \cline{1-3}
    Times & 183s & 154s \\ \cline{1-3}
  \end{tabular}
  \caption{\footnotesize MAE and execution times for the deconvolution of the sky image.}
  \label{tab:maesky}
\end{table}

\subsection{Inpainting with Gaussian noise}
\label{sec:inpa-with-gauss}

\begin{figure}[htb]
\hspace*{-0.5cm}
  \centering
  \scriptsize
  \begin{tabular}{c@{ }c}
    \includegraphics[width=0.4\linewidth]{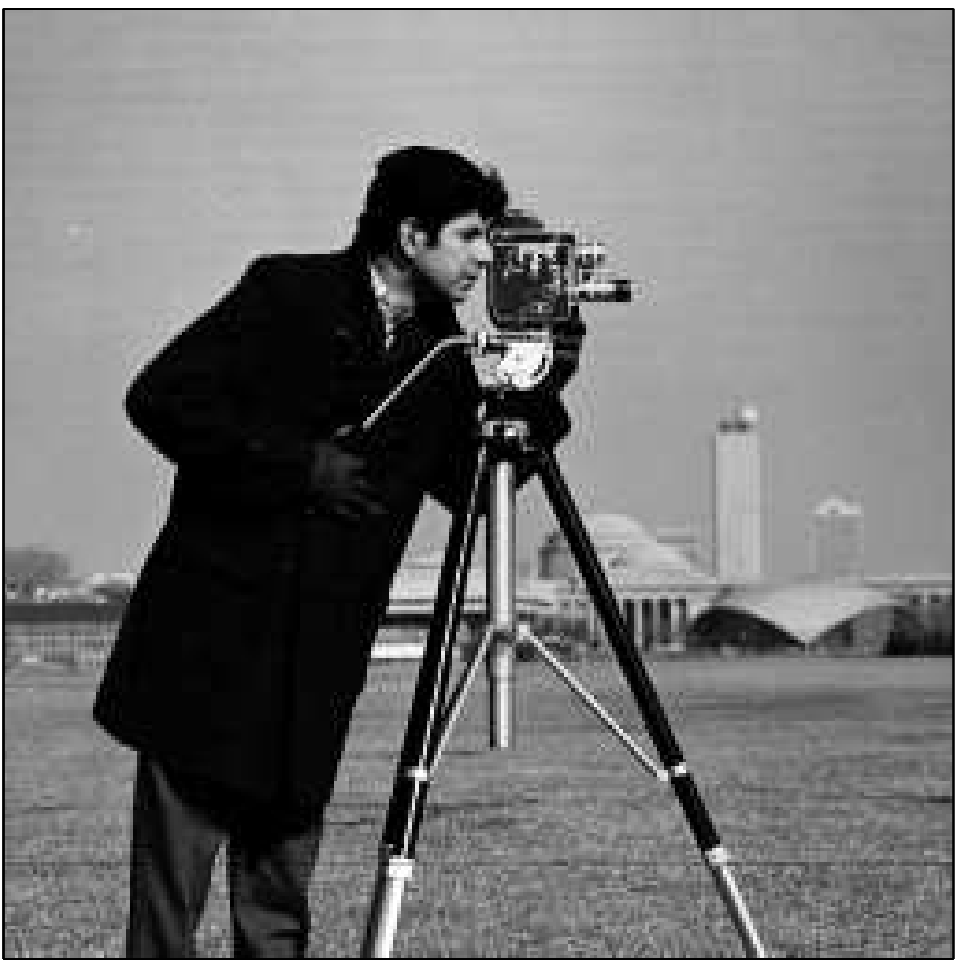} &
    \includegraphics[width=0.4\linewidth]{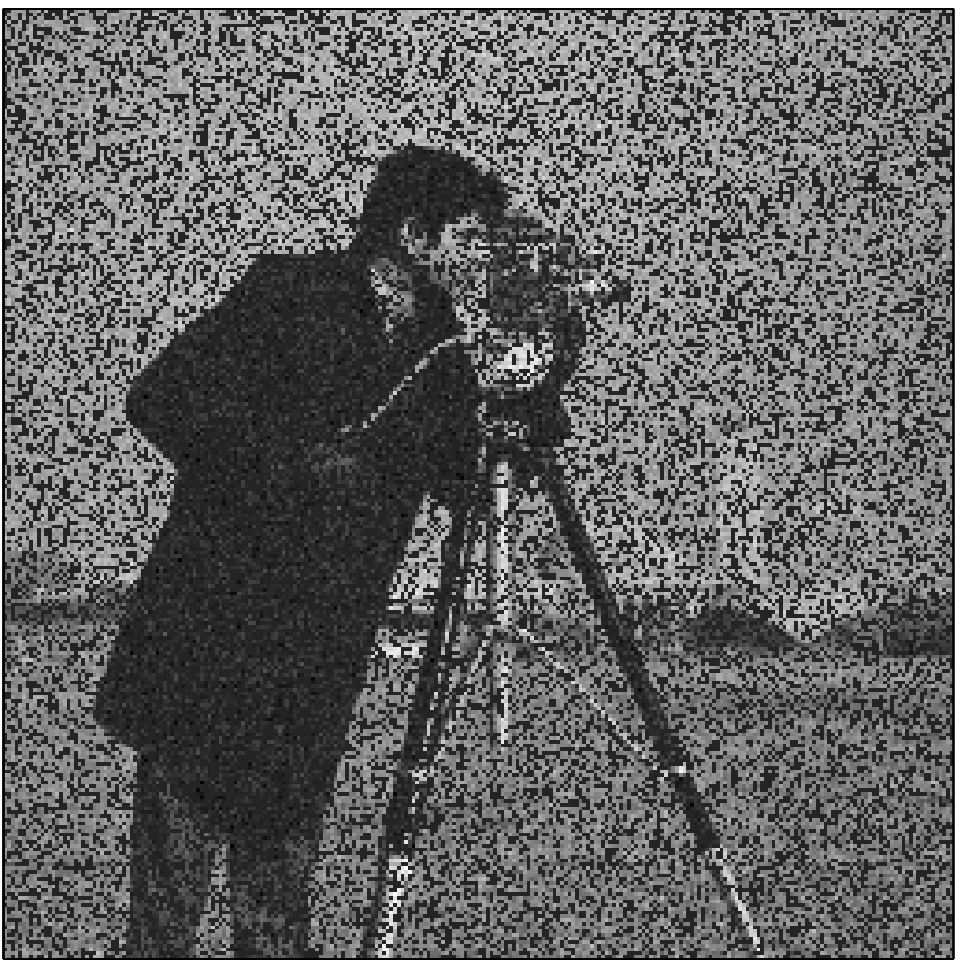} \\
    Original & Masked and noisy (PSNR = 11.1) \\
    \includegraphics[width=0.4\linewidth]{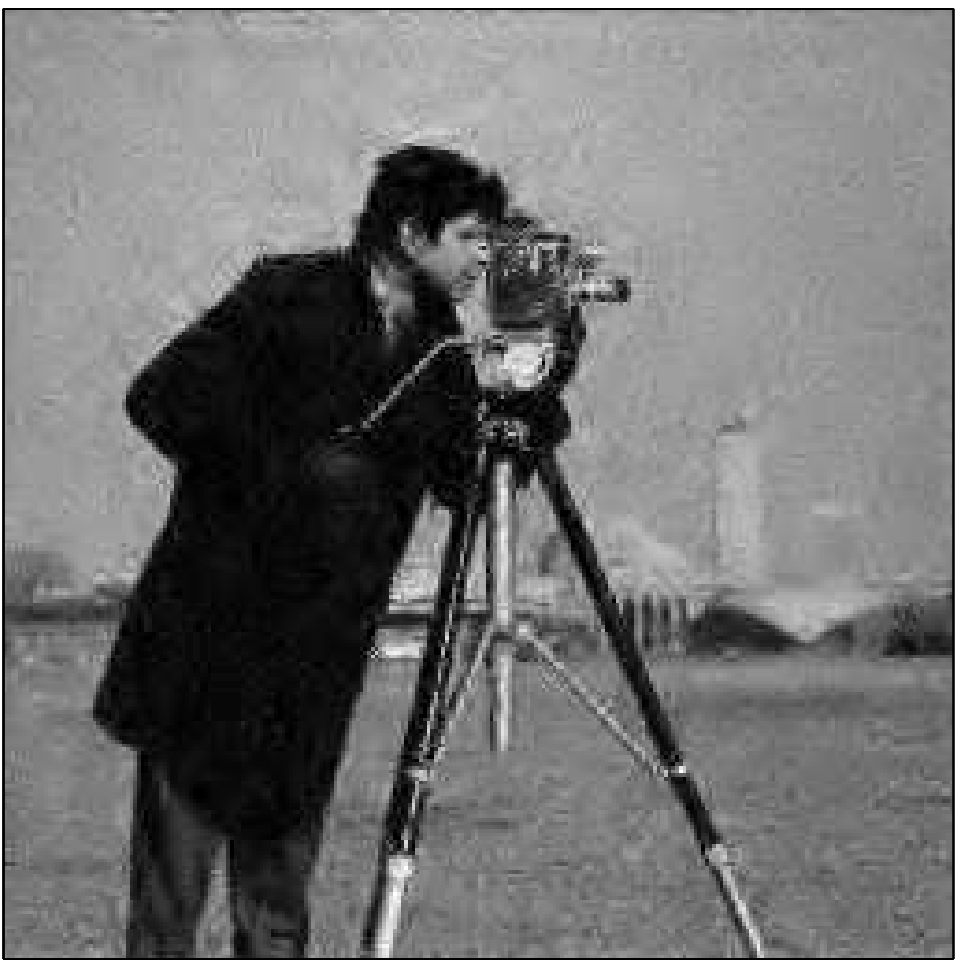} &
    \includegraphics[width=0.4\linewidth]{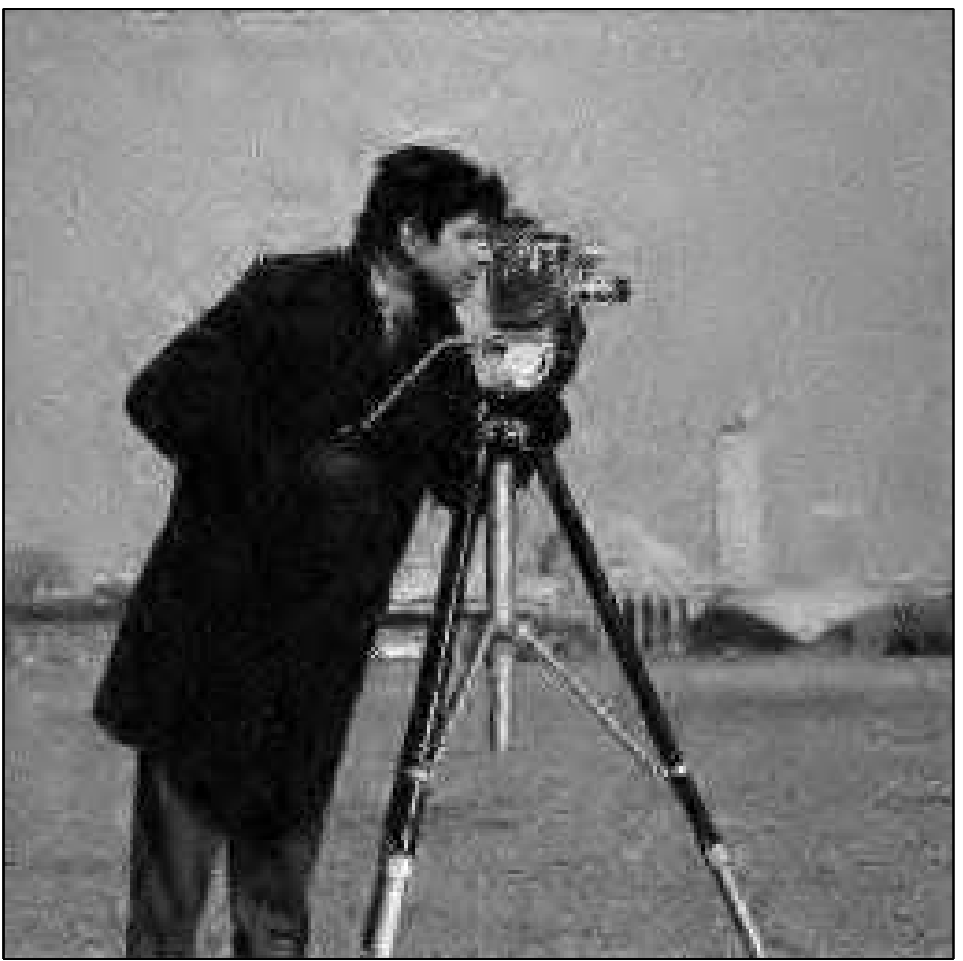} \\
    Alg.~\ref{algo:deconv} (PSNR = 25.8) & Alg.~\ref{algo:deconv2} (PSNR = 25.8) \\
  \end{tabular}
  \caption{\footnotesize Inpainting results for the Cameraman using our two algorithms.}
  \label{fig:psnrcam}
\end{figure}

\begin{figure}[htb]
  \centering
  \includegraphics[width=0.4\linewidth]{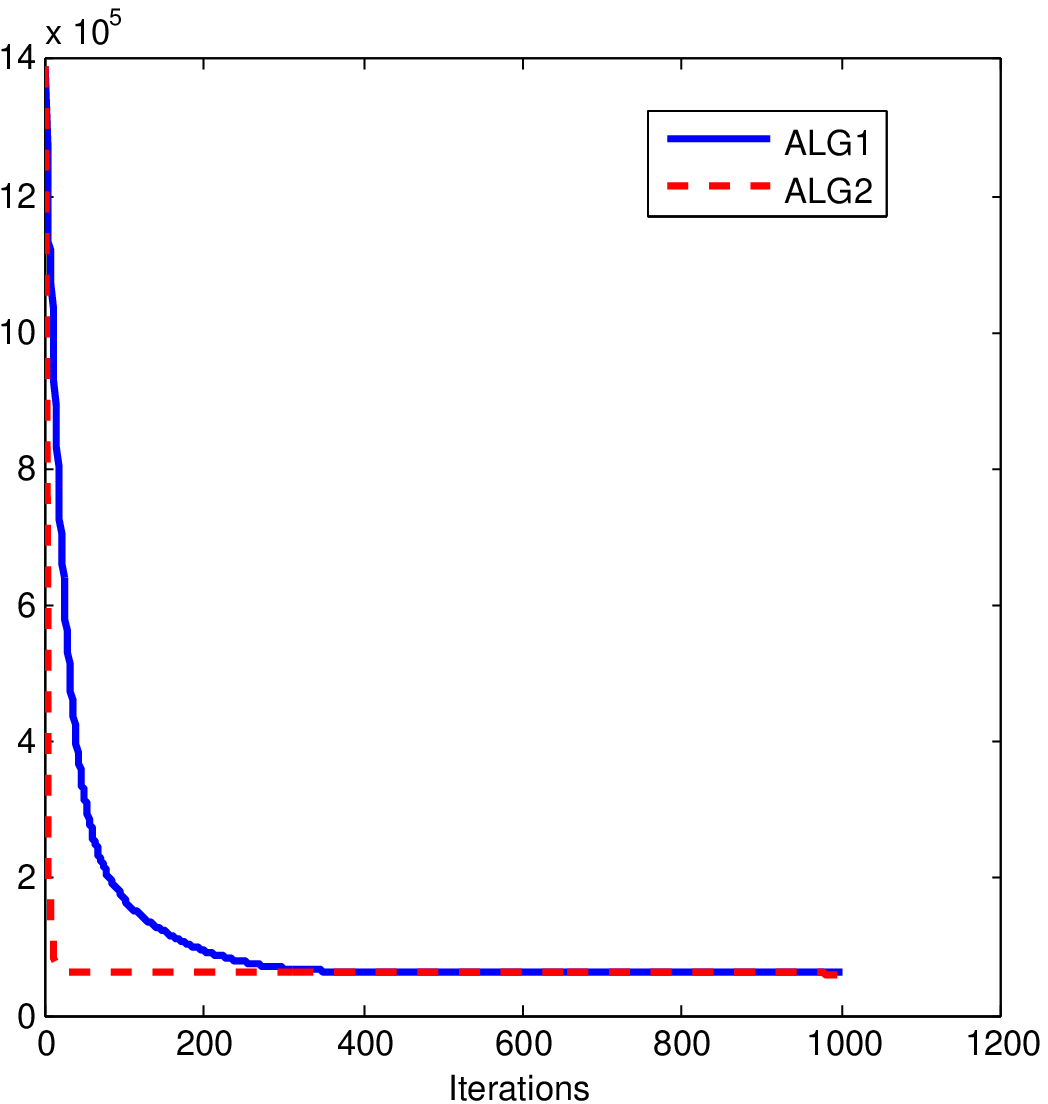}
  \includegraphics[width=0.4\linewidth]{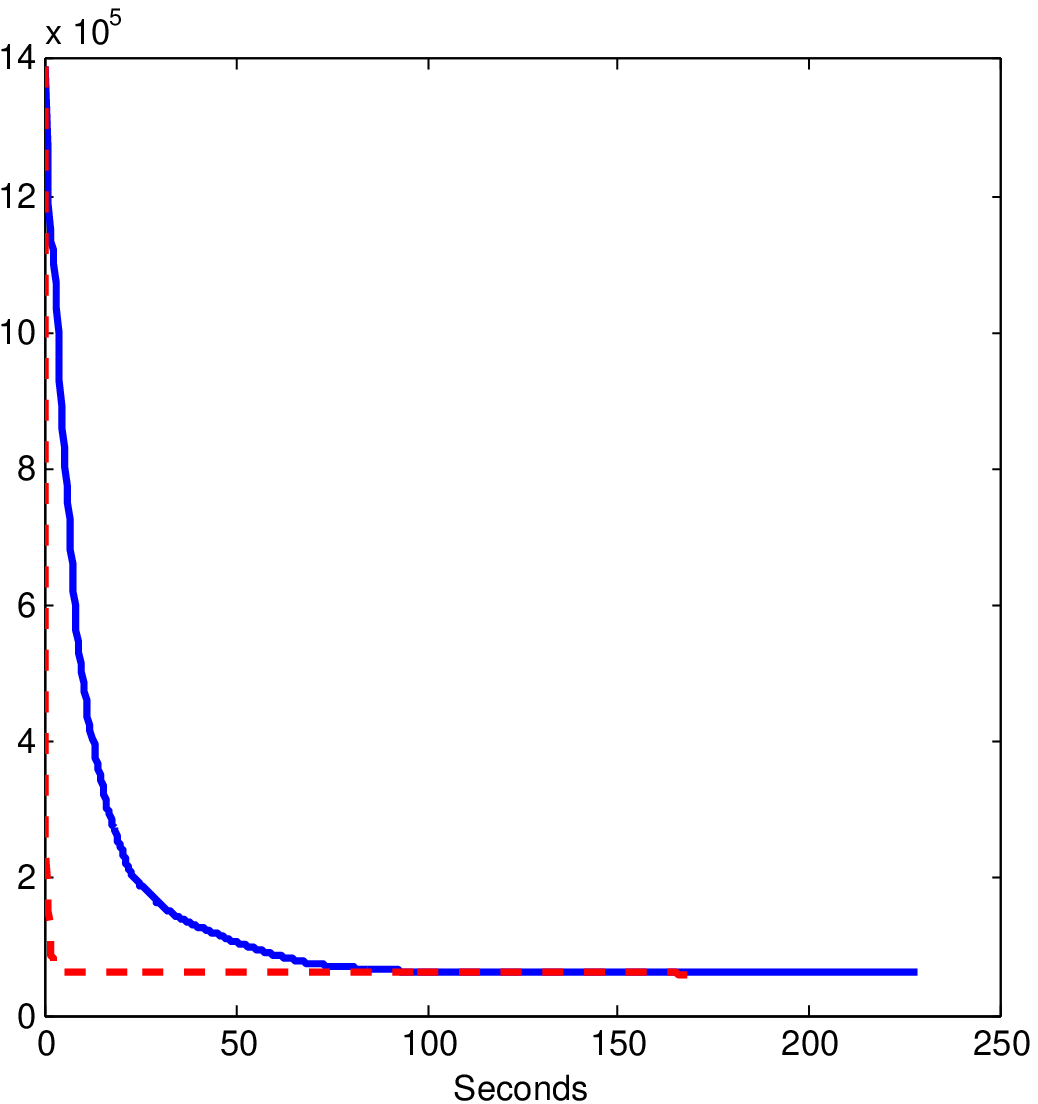}
  \caption{Objective function for inpainting with Gaussian noise in function if iterations (left) and times (right).}
  \label{fig:plot2}
\end{figure}

We also applied our algorithms to inpainting with Gaussian noise.  In all experiments, $\Psi$ was the
$\ell_1$-norm. Fig~\ref{fig:psnrcam} summarizes the results with the PSNR and the execution times for the Cameraman,
where the dictionary consisted of the wavelet transform and the mask was create from a random process (here with about
34\% of missing pixels). Notice that both algorithms leads to the same solution which gives a good reconstruction of the
image.  Fig.~\ref{fig:plot2} displays the objective as a function of the iteration number and times (in s). Again, we
can clearly see that Algorithm~2 converges faster than Algorithm~1.

\subsection{Denoising with Multiplicative noise}
\label{sec:inpa-with-mult}

\begin{figure}[htb]
\hspace*{-0.5cm}
  \centering
  \scriptsize
  \begin{tabular}{c@{ }c}
    \includegraphics[width=0.4\linewidth]{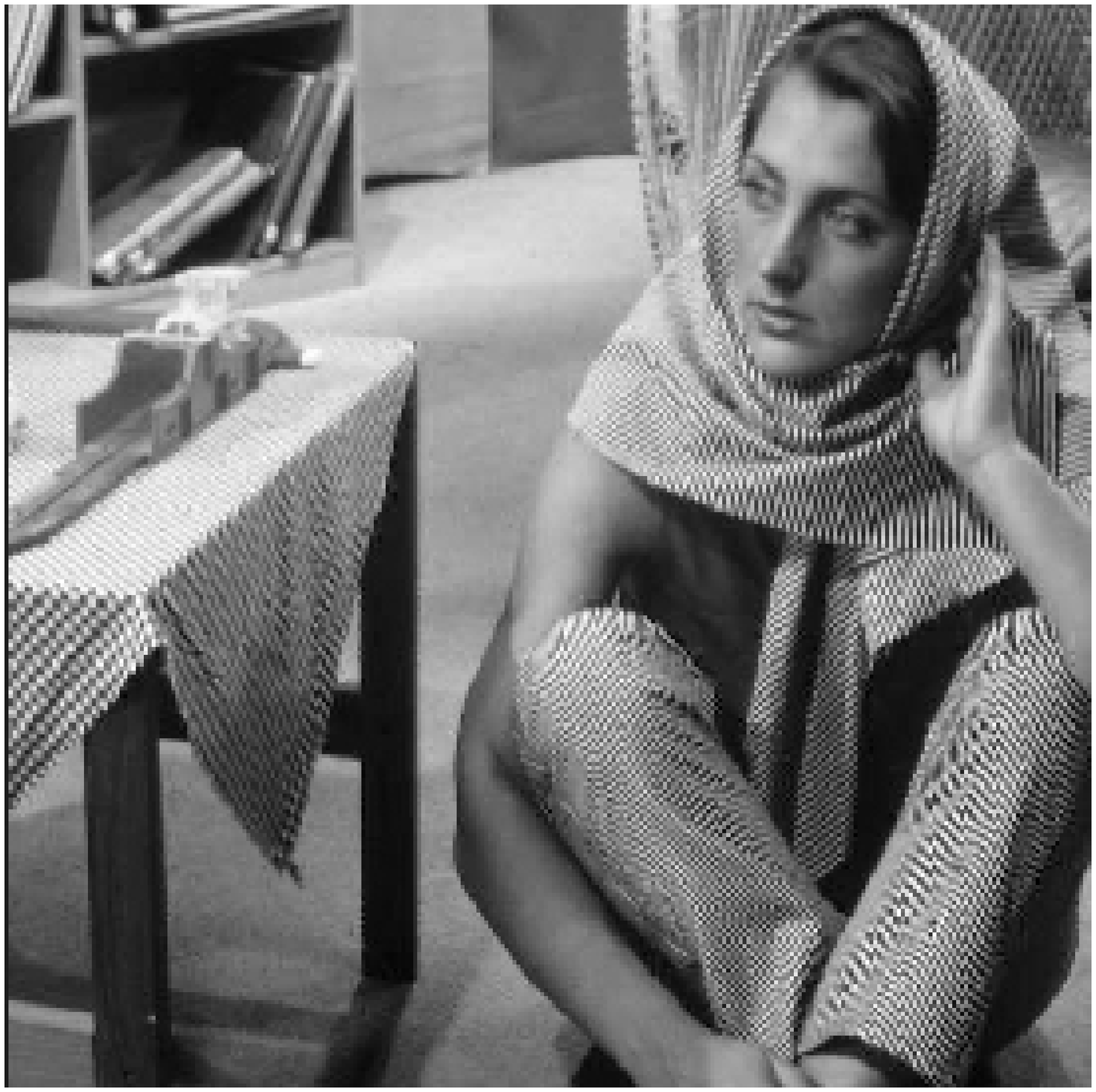} &
    \includegraphics[width=0.4\linewidth]{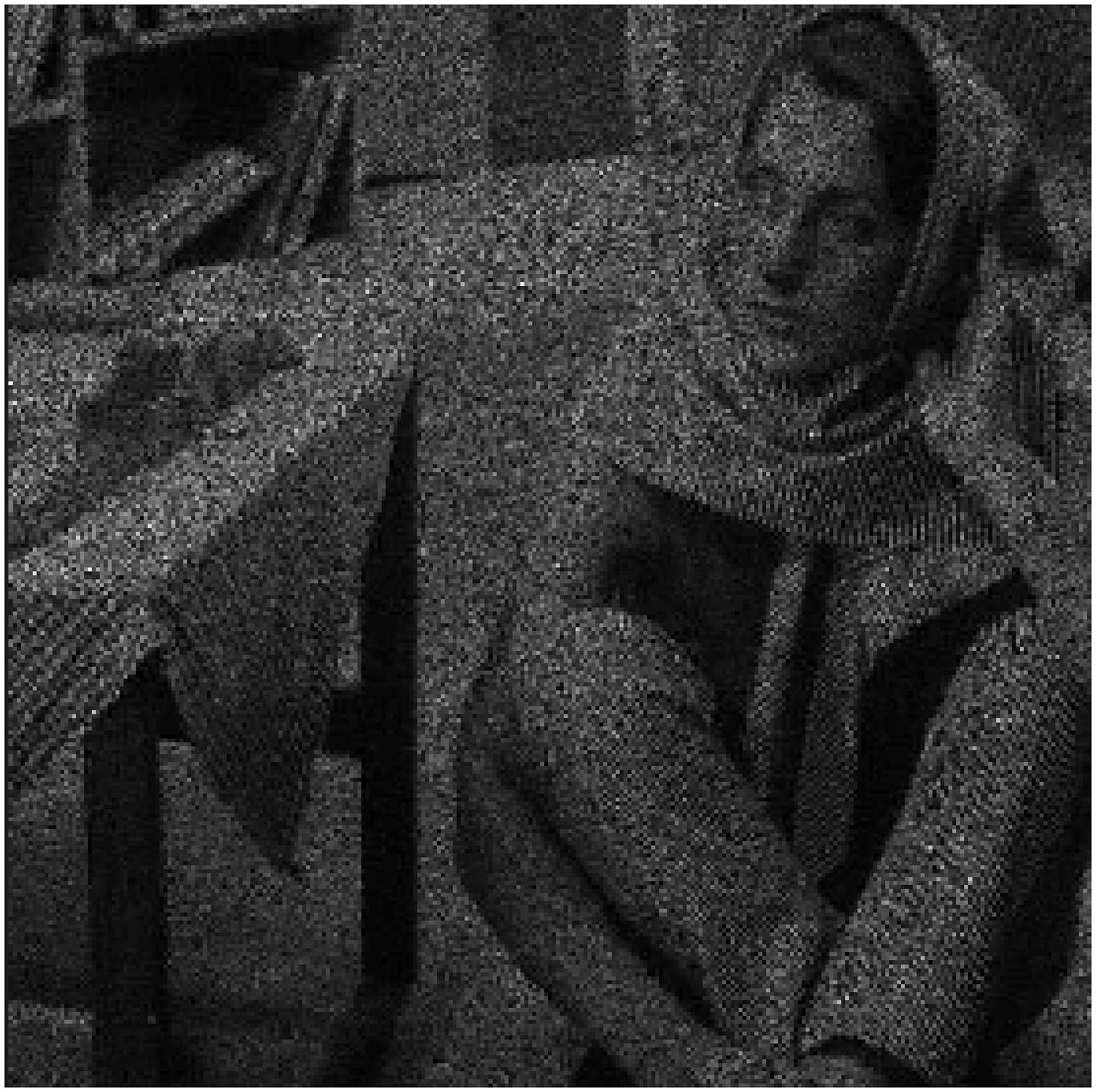} \\
    Original & Masked and noisy (MAE = 3.6) \\
    \includegraphics[width=0.4\linewidth]{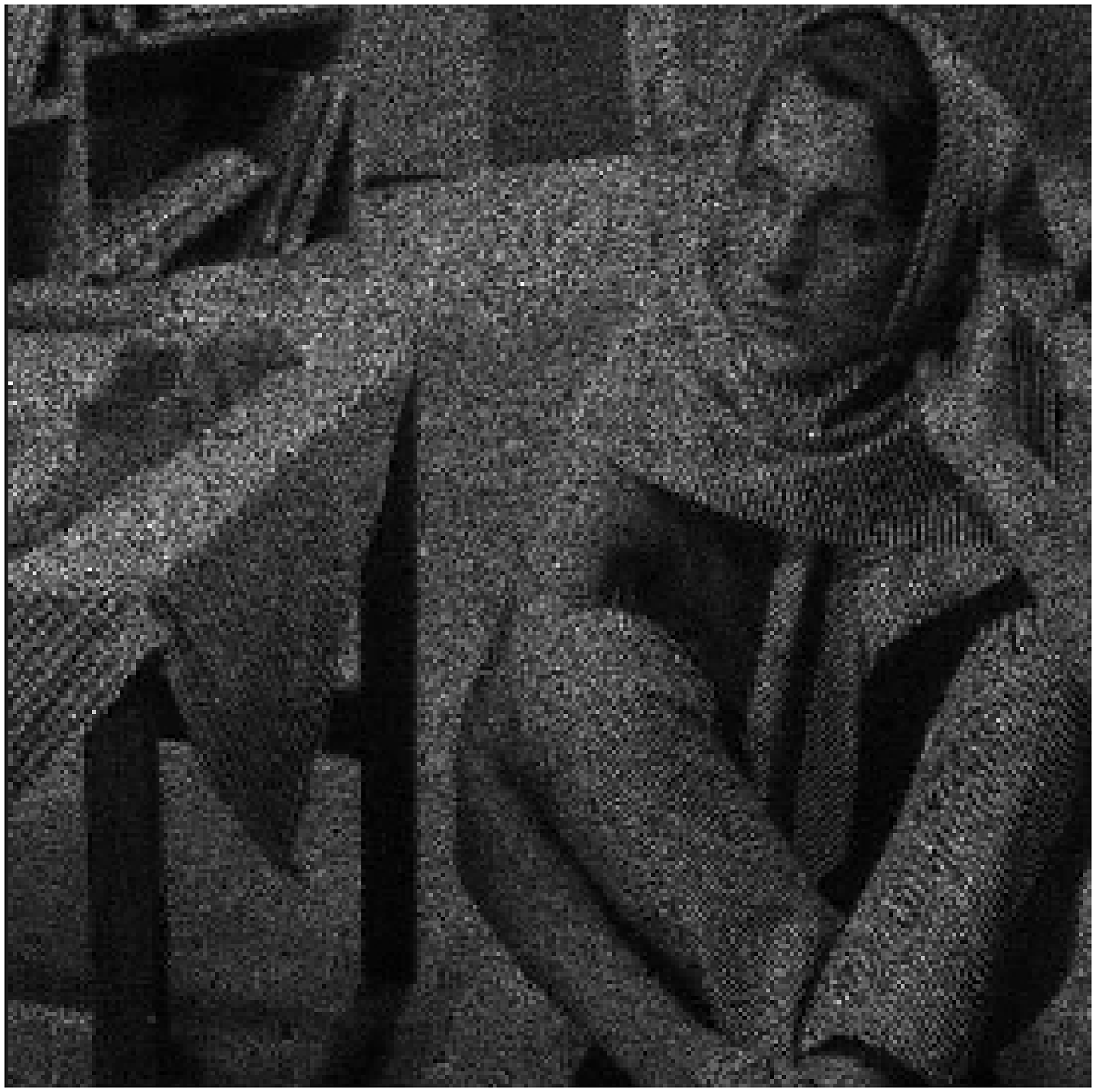} &
    \includegraphics[width=0.4\linewidth]{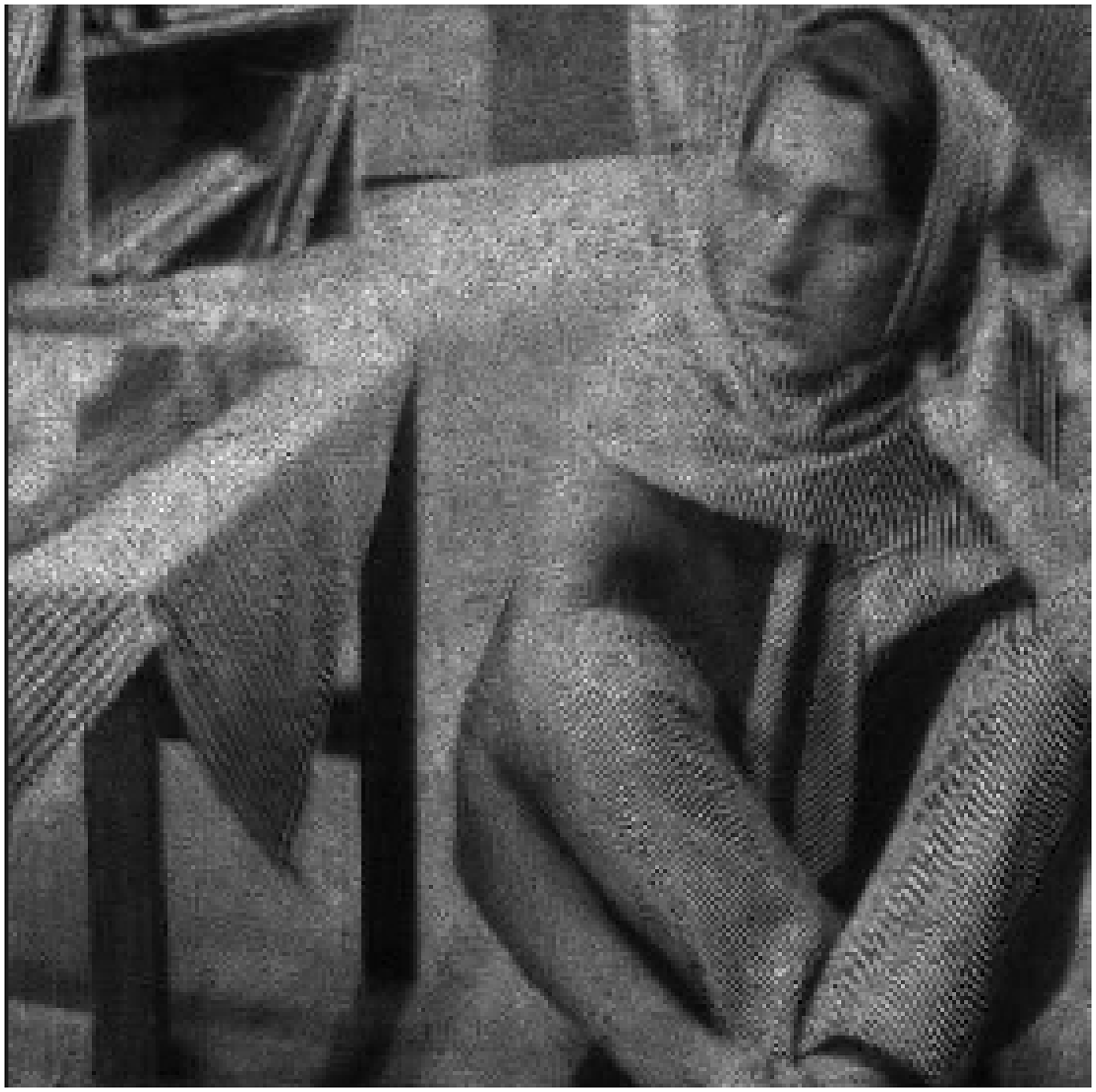} \\
    Alg.~\ref{algo:deconv} (MAE = 3.2) & Alg.~\ref{algo:deconv2} (MAE = 2.3) \\
  \end{tabular}
  \caption{\footnotesize Denoising results for Barbara using our two algorithms.}
  \label{fig:psnrbar}
\end{figure}

\begin{figure}[htb]
  \centering
  \includegraphics[width=0.4\linewidth]{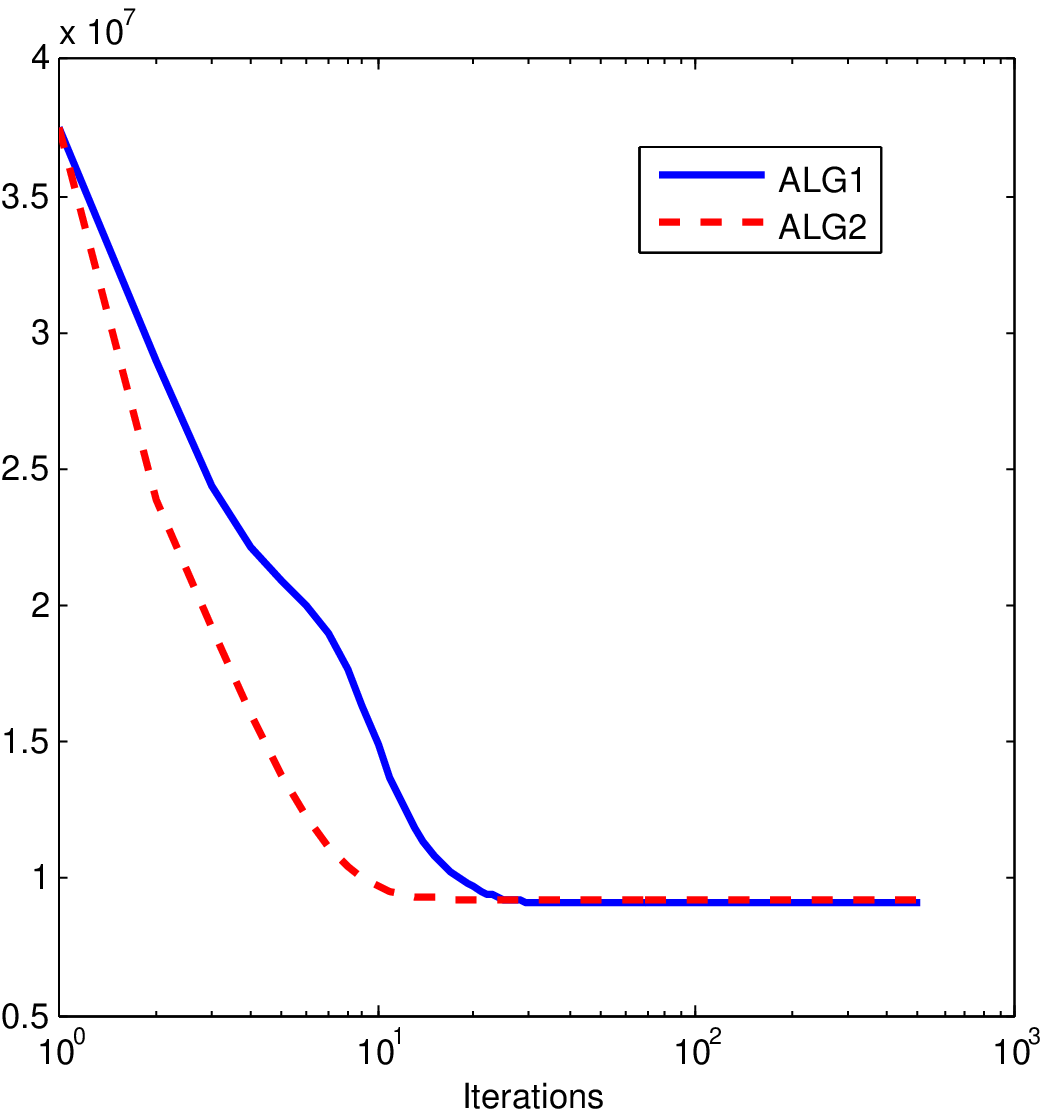}
  \includegraphics[width=0.4\linewidth]{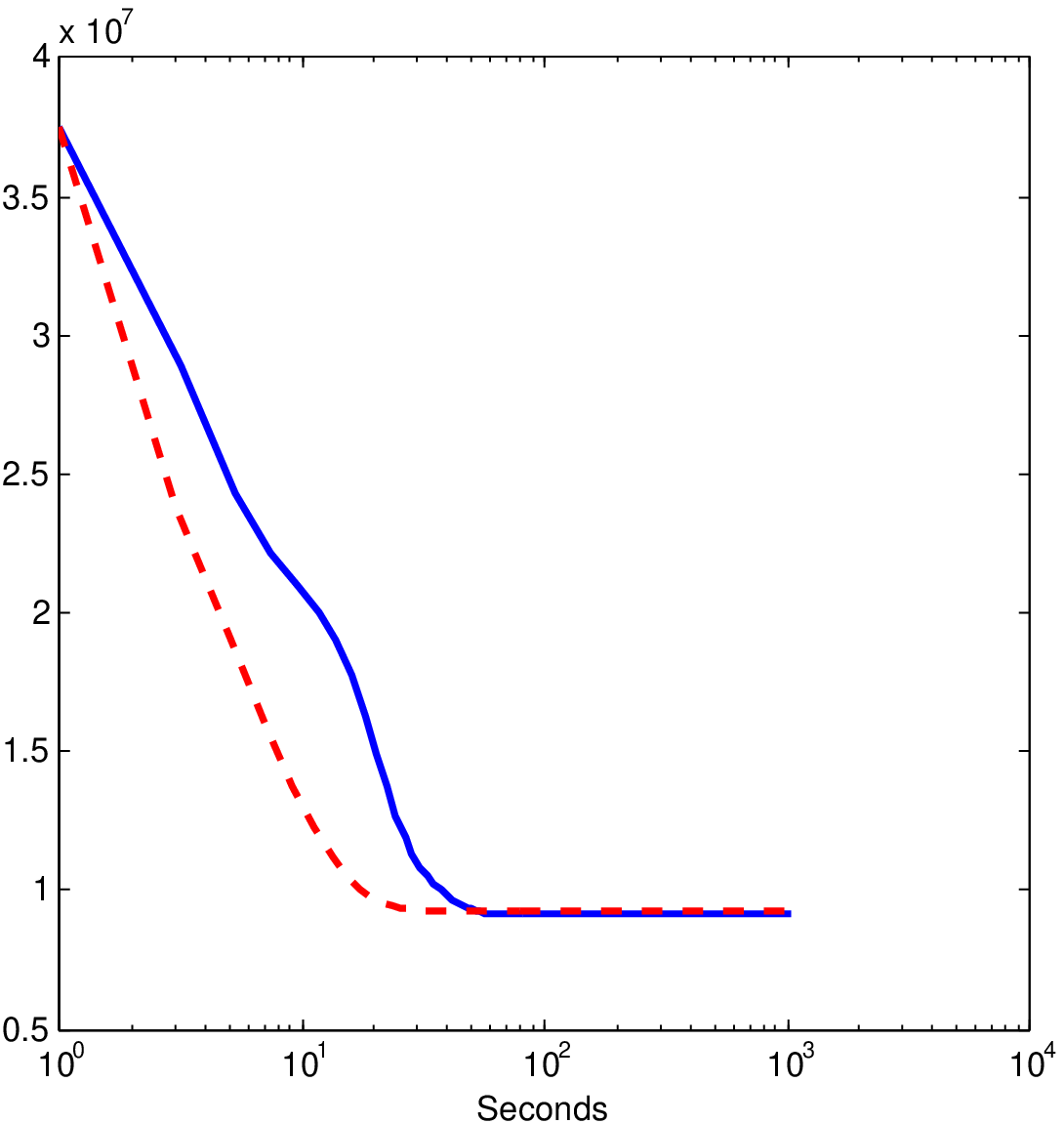}
  \caption{Objective function for denoising with multiplicative noise in function if iterations (left) and times (right).}
  \label{fig:plot3}
\end{figure}

As we work on the logarithm the problem (see~\ref{sec:multiplicative-noise}, the final estimate for each algorithm is
given by taking the exponential of the result. In all experiments, $\Psi$ was the $\ell_1$-norm. The Barbara image was
set to a maximal intensity of 30 and the minimal to a non-zero value in order to avoid issues with the logarithm. The
noise was added using $M=10$ which leads to a medium level of noise.  Fig~\ref{fig:psnrbar} summarizes the results with
the MAE and the execution times for Barbara, where the dictionary consisted of the curvelets transform. Our methods give
correct reconstruction of the image. Fig.~\ref{fig:plot3} displays the objective as a function of the iteration number
and times (in s). Again, we can clearly see that Algorithm~2 converges faster than Algorithm~1.

\section{Conclusion}
\label{sec:conclusion}

In this paper, we proposed two provably convergent algorithms for solving the linear inverse problems with a sparsity
prior. The primal-dual proximal splitting algorithm seems to perform better in terms of convergence speed than the
primal one. Moreover, its computational burden is lower than most comparable of state-of-art methods. Inverse problems
with multiplicative noise does not enter currently in this framework, we will consider its adaptation to such problems
in future work.

%
\bibliographystyle{abbrv}
\bibliography{references}  
%
%

\end{document}

%% file: macros.tex
\newcommand{\eqi}{\gamma}

\newcommand{\kerm}{\mathrm{ker}}

\newcommand{\va}{\alpha}
\newcommand{\vx}{{x}}
\newcommand{\vy}{{y}}
\newcommand{\parenth}[1]{\left({#1}\right)}

\newcommand{\Pm}{\mathrm{P}}

\newcommand{\Id}{\boldsymbol{\mathrm{I}}}

\newcommand{\Hmb}{\boldsymbol{\mathrm{H}}}
\newcommand{\Kmb}{\boldsymbol{\mathrm{K}}}
\newcommand{\Lmb}{\boldsymbol{\mathrm{L}}}

\newcommand{\Phb}{\boldsymbol{\Phi}}

\newcommand{\opnorm}[1]{\left\lvert\!\left\lvert\!\left\lvert #1 \right\rvert\!\right\rvert\!\right\rvert}
\newcommand{\norm}[1]{\left\lVert #1 \right\rVert}

\newcommand{\abs}[1]{\left\lvert #1 \right\rvert}

\newcommand{\Cc}{\mathcal{C}}

\newcommand{\Hc}{\mathcal{H}}

\newcommand{\Kc}{\mathcal{K}}

\newcommand{\Mc}{\mathcal{M}}

\newcommand{\Pc}{\mathcal{P}}

\newcommand{\Prj}{\mathcal{P}}

\DeclareMathOperator{\prox}{prox}

\DeclareMathOperator{\dom}{dom}
\DeclareMathOperator{\ri}{ri}

\DeclareMathOperator*{\argmin}{argmin}

\renewcommand{\ge}{\geqslant}
\renewcommand{\le}{\leqslant}

\newtheorem{theorem}{\textbf{Theorem}} 

\newtheorem{lemma}[theorem]{\textbf{Lemma}}

\newtheorem{propo}[theorem]{\textbf{Proposition}}

\newtheorem{definition}[theorem]{\textbf{Definition}}

